\documentclass
[prl,10pt,letterpaper,draft,bibnotes,notitlepage,final,superscriptaddress,balancelastpage]{revtex4}
\usepackage{epsfig}
\usepackage{amsmath}
\usepackage{graphicx}
\usepackage{bm}

\begin{document}

\title{An experimental test of non-local realism}
\author{Simon Gr\"oblacher}\affiliation{Faculty of Physics, University of Vienna, Boltzmanngasse 5, A-1090 Vienna, Austria}\affiliation{Institute for Quantum Optics and Quantum Information (IQOQI), Austrian Academy of
Sciences, Boltzmanngasse 3, A-1090 Vienna, Austria}
\author{Tomasz Paterek}\affiliation{Institute of Theoretical Physics and Astrophysics, University of Gdansk, ul.\ Wita Stwosza 57, PL-08-952 Gdansk, Poland}\affiliation{The Erwin Schr\"odinger International Institute for Mathematical Physics (ESI), Boltzmanngasse 9, A-1090 Vienna, Austria}
\author{Rainer Kaltenbaek}\affiliation{Faculty of Physics, University of Vienna, Boltzmanngasse 5, A-1090 Vienna, Austria}
\author{\v{C}aslav Brukner}\affiliation{Faculty of Physics, University of Vienna, Boltzmanngasse 5, A-1090 Vienna, Austria}\affiliation{Institute for Quantum Optics and Quantum Information (IQOQI), Austrian Academy of
Sciences, Boltzmanngasse 3, A-1090 Vienna, Austria}
\author{Marek \.Zukowski}\affiliation{Faculty of Physics, University of Vienna, Boltzmanngasse 5, A-1090 Vienna, Austria}\affiliation{Institute of Theoretical Physics and Astrophysics, University of Gdansk, ul.\ Wita Stwosza 57, PL-08-952 Gdansk, Poland}
\author{Markus Aspelmeyer}\email{markus.aspelmeyer@quantum.at}\affiliation{Faculty of Physics, University of Vienna, Boltzmanngasse 5, A-1090 Vienna, Austria}\affiliation{Institute for Quantum Optics and Quantum Information (IQOQI), Austrian Academy of
Sciences, Boltzmanngasse 3, A-1090 Vienna, Austria}
\author{Anton Zeilinger}\email{zeilinger-office@quantum.at}\affiliation{Faculty of Physics, University of Vienna, Boltzmanngasse 5, A-1090 Vienna, Austria}\affiliation{Institute for Quantum Optics and Quantum Information (IQOQI), Austrian Academy of
Sciences, Boltzmanngasse 3, A-1090 Vienna, Austria}

\begin{abstract}
Most working scientists hold fast to the concept of 'realism' - a viewpoint according to which an external reality exists independent of observation. But quantum physics has shattered some of our cornerstone beliefs. According to Bell's theorem, any theory that is based on the joint assumption of realism and locality (meaning that local events cannot be affected by actions in space-like separated regions) is at variance with certain quantum predictions. Experiments with entangled pairs of particles have amply confirmed these quantum predictions, thus rendering local realistic theories untenable. Maintaining realism as a fundamental concept would therefore necessitate the introduction of 'spooky' actions that defy locality. Here we show by both theory and experiment that a broad and rather reasonable class of such non-local realistic theories is incompatible with experimentally observable quantum correlations. In the experiment, we measure previously untested correlations between two entangled photons, and show that these correlations violate an inequality proposed by Leggett for non-local realistic theories. Our result suggests that giving up the concept of locality is not sufficient to be consistent with quantum experiments, unless certain intuitive features of realism are abandoned~\cite{Groeblacher2007}.
\end{abstract}

\maketitle

Physical realism suggests that the results of observations are a consequence of properties carried by physical systems. It remains surprising that this tenet is very little challenged, as its significance goes far beyond science. Quantum physics, however, questions this concept in a very deep way. To maintain a realistic description of nature, non-local hidden-variable theories are being discussed as a possible completion of quantum theory. They offer to explain intrinsic quantum phenomena - above all, quantum entanglement~\cite{Schroedinger1935} - by nonlocal influences. Up to now, however, it has not been possible to test such theories in experiments. We present an inequality, similar in
spirit to the seminal one given by Clauser, Horne, Shimony and Holt~\cite{CHSH1969} on local hidden variables, that allows us to test an important class of non-local hidden-variable theories against quantum theory. The theories under test provide an explanation of all existing two qubit Bell-type experiments. Our derivation is based on a recent incompatibility theorem by Leggett~\cite{Leggett2003}, which we extend so as to make it applicable to real experimental situations and also to allow simultaneous tests of all local hidden-variable models. Finally, we perform an experiment that violates the new inequality and hence excludes for the first time a broad class of non-local hidden-variable theories.\\

Quantum theory gives only probabilistic predictions for individual events. Can one go beyond this? Einstein's view~\cite{EPR1935,Einstein1935} was that quantum theory does not provide a complete description of physical reality: ''While we have thus shown that the wavefunction does not provide a complete description of the physical reality, we left open the question of whether or not such a description exists. We believe, however, that such a theory is possible.''~\cite{EPR1935}. It remained an open question whether the theory could be completed in Einstein's sense~\cite{Bohr1949}. If so, more complete theories based on objective properties of physical systems should be possible. Such models are referred to as hidden-variable theories.

Bell's theorem~\cite{Bell1964} proves that all hidden-variable theories based on the joint assumption of locality and realism are at variance with the predictions of quantum physics. Locality prohibits any influences between events in space-like separated regions, while realism claims that all measurement outcomes depend on pre-existing properties of objects that are independent of the measurement. The more refined versions of Bell's theorem by Clauser, Horne, Shimony and Holt~\cite{CHSH1969} and by Clauser and Horne~\cite{CH1974,Clauser2002} start from the assumptions of local realism and result in inequalities for a set of statistical correlations (expectation values), which must be satisfied by all local realistic hidden-variable theories. The inequalities are violated by quantum mechanical predictions. Greenberger, Horne and Zeilinger~\cite{GHZ1989,Greenberger1990} showed that already perfect correlations of systems with at least three particles are inconsistent with these assumptions. So far, all experiments motivated by these theorems are in full agreement with quantum predictions~\cite{Freedman1972,Aspect1981,Aspect1982,Weihs1998,Rowe2001,Pan2000}. For some time, loopholes existed that allowed the observed correlations to be explained within local realistic theories. In particular, an ideal Bell experiment has to be performed with detectors of sufficiently high efficiency (to close the 'detection loophole') and with experimental settings that are randomly chosen in space-like separated regions (to close the 'locality loophole'). Since the first successful Bell experiment by Freedman and Clauser~\cite{Freedman1972}, later implementations have continuously converged to closing both the locality loophole~\cite{Aspect1982,Weihs1998,Zeilinger1986,Aspect1999} on the one hand and the detection loophole~\cite{Rowe2001,Grangier2001} on the other hand. Therefore it is reasonable to consider the violation of local realism a well established fact.\\

The logical conclusion one can draw from the violation of local realism is that at least one of its assumptions fails. Specifically, either locality or realism or both cannot provide a foundational basis for quantum theory. Each of the resulting possible positions has strong supporters and opponents in the scientific community. However, Bell's theorem is unbiased with respect to these views: on the basis of this theorem, one cannot, even in principle, favour one over the other. It is therefore important to ask whether incompatibility theorems similar to Bell's can be found in which at least one of these concepts is relaxed. Our work addresses a broad class of non-local hidden-variable theories that are based on a very plausible type of realism and that provide an explanation for all existing Bell-type experiments. Nevertheless we demonstrate, both in theory and experiment, their conflict with quantum predictions and observed measurement data. Following the recent approach of Leggett~\cite{Leggett2003}, who introduced the class of non-local models and formulated an incompatibility theorem, we have analysed its assumptions and derived an inequality valid for such theories that can be experimentally tested. In addition, the experiments allow for a simultaneous test of all local hidden-variable models - that is, the measurement data can neither be explained by a local realistic model nor by the considered class of non-local models.\\

The theories under investigation describe experiments on pairs of particles. It is sufficient for our purposes to discuss two-dimensional quantum systems. We will hence focus our description on the polarization degree of freedom of photons. The theories are based on the following assumptions: (1) all measurement outcomes are determined by pre-existing properties of particles independent of the measurement (realism); (2) physical states are statistical mixtures of subensembles with definite polarization, where (3) polarization is defined such that expectation values taken for each subensemble obey Malus' law (that is, the well-known cosine dependence of the intensity of a polarized beam after an ideal polarizer).\\

These assumptions are in a way appealing, because they provide a natural explanation of quantum mechanically separable states (polarization states indeed obey Malus' law). In addition, they do not explicitly demand locality; that is, measurement outcomes may very well depend on parameters in space-like separated regions. As a consequence, such theories can explain important features of quantum mechanically entangled (non-separable) states of two particles (a specific model can be found in Appendix I):\ first, they do not allow information to be transmitted faster than the speed of light; second, they reproduce perfect correlations for all measurements in the same bases, which is a fundamental feature of the Bell singlet state; and third, they provide a model for all thus far performed experiments in which the Clauser, Horne, Shimony and Holt (CHSH) inequality was violated. Nevertheless, we will show that all models based on assumptions (1)-(3) are at variance with other quantum predictions.

A general framework of such models is the following: assumption (1) requires that an individual binary measurement outcome $A$ for a polarization measurement along direction $\vec{a}$ (that is, whether a single photon is transmitted or absorbed by a polarizer set at a specific angle) is predetermined by some set of hidden-variables $\lambda$, and a three-dimensional vector $\vec{u}$, as well as by some set of other possibly non-local parameters $\eta$ (for example, measurement settings in space-like separated regions) - that is, $A=A(\lambda,\vec{u},\vec{a},\eta)$. According to assumption (3), particles with the same $\vec{u}$ but with different $\lambda$ build up subensembles of 'definite polarization' described by a probability distribution $\rho_{\vec{u}}(\lambda)$. The expectation value $\overline{A}(\vec{u})$, obtained by averaging over $\lambda$, fulfils Malus' law, that is, $\overline{A}(\vec{u})=\int d\lambda\rho_{\vec{u}}(\lambda)A(\lambda,\vec{u},\vec{a},\eta)=\vec{u}\cdot\vec{a}$. Finally, with assumption (2), the measured expectation value for a general physical state is given by averaging over the distribution $F(\vec{u})$ of subensembles, that is, $\langle A\rangle=\int d\vec{u}F(\vec{u})\overline{A}(\vec{u})$.\\

Let us consider a specific source, which emits pairs of photons with well-defined polarizations $\vec{u}$ and $\vec{v}$ to laboratories of Alice and Bob, respectively. The local polarization measurement outcomes $A$ and $B$ are fully determined by the polarization vector, by an additional set of hidden variables $\lambda$ specific to the source and by any set of parameters $\eta$ outside the source. For reasons of clarity, we choose an explicit non-local dependence of the outcomes on the settings $\vec{a}$ and $\vec{b}$ of the measurement devices. Note, however, that this is just an example of a possible non-local dependence, and that one can choose any other set out of $\eta$. Each emitted pair is fully defined by the subensemble distribution $\rho_{\vec{u},\vec{v}}(\lambda)$. In agreement with assumption (3) we impose the following conditions on the predictions for local averages of such measurements (all polarizations and measurement directions are represented as vectors on the Poincar\'{e} sphere~\cite{Born1964}):
\begin{eqnarray}
\overline{A}(\vec{u})=\int d\lambda\rho_{\vec{u},\vec{v}}(\lambda)A(\vec{a},\vec{b},\lambda)=\vec{u}\cdot\vec{a}\label{eq1}\\
\overline{B}(\vec{v})=\int d\lambda\rho_{\vec{u},\vec{v}}(\lambda)B(\vec{b},\vec{a},\lambda)=\vec{v}\cdot\vec{b}\label{eq2}
\end{eqnarray}
It is important to note that the validity of Malus' law imposes the non-signalling condition on the investigated non-local models, as the local expectation values do only depend on local parameters. The correlation function of measurement results for a source emitting well-polarized photons is defined as the average of the products of the individual measurement outcomes:
\begin{equation}
\overline{AB}(\vec{u},\vec{v})=\int d\lambda\rho_{\vec{u},\vec{v}}(\lambda)A(\vec{a},\vec{b},\lambda)B(\vec{b},\vec{a},\lambda)
\end{equation}
For a general source producing mixtures of polarized photons the observable correlations are averaged over a distribution of the polarizations $F(\vec{u},\vec{v})$, and the general correlation function $E$ is given by:
\begin{equation}
E=\langle AB\rangle=\int d\vec{u}d\vec{v}F(\vec{u},\vec{v})\overline{AB}(\vec{u},\vec{v})\label{eq4}
\end{equation}
It is a very important trait of this model that there exist subensembles of definite polarizations (independent of measurements) and that the predictions for the subensembles agree with Malus' law. It is clear that other classes of non-local theories, possibly even fully compliant with all quantum mechanical predictions, might exist that do not have this property when reproducing entangled states. Such theories may, for example, include additional communication~\cite{Bacon2003} or dimensions~\cite{Ne'eman1986}. A specific case deserving comment is Bohm's theory~\cite{Bohm1952}. There the non-local correlations are a consequence of the non-local quantum potential, which exerts suitable torque on the particles leading to experimental results compliant with quantum mechanics. In that theory, neither of the two particles in a maximally entangled state carries any angular momentum at all when emerging from the source~\cite{Dewdney1987}. In contrast, in the Leggett model, it is the total ensemble emitted by the source that carries no angular momentum, which is a consequence of averaging over the individual particles' well defined angular momenta (polarization).\\

The theories described here are incompatible with quantum theory. The basic idea of the incompatibility theorem~\cite{Leggett2003} uses the following identity, which holds for any numbers $A=\pm1$ and $B=\pm1$:
\begin{equation}
-1+|A+B|=AB=1-|A-B|
\end{equation}
One can apply this identity to the dichotomic measurement results $A=A(\vec{a},\vec{b},\lambda)=\pm1$ and $B=B(\vec{b},\vec{a},\lambda)=\pm1$. The identity holds even if the values of $A$ and $B$ mutually depend on each other. For example, the value of a specific outcome $A$ can depend on the value of an actually obtained result $B$. In contrast, in the derivation of the CHSH inequality it is necessary to assume that $A$ and $B$ do not depend on each other. Therefore, any kind of non-local dependencies used in the present class of theories are allowed. Taking the average over the subensembles with definite polarizations we obtain:
\begin{equation}
-1+\int d\lambda\rho_{\vec{u},\vec{v}}(\lambda)|A+B|=\int d\lambda\rho_{\vec{u},\vec{v}}(\lambda)AB=1-\int d\lambda\rho_{\vec{u},\vec{v}}(\lambda)|A-B|
\end{equation}
Denoting these averages by bars, one arrives at the shorter expression:
\begin{equation}
-1+\overline{|A+B|}=\overline{AB}=1-\overline{|A-B|}
\end{equation}
As the average of the modulus is greater than or equal to the modulus of the averages, one gets the set of inequalities:
\begin{equation}
-1+|\overline{A}+\overline{B}|\leq\overline{AB}\leq1-|\overline{A}-\overline{B}|\label{eq8}
\end{equation}
By inserting Malus' law, equations (\ref{eq1}) and (\ref{eq2}), in equation (\ref{eq8}), and by using expression (\ref{eq4}), one arrives at a set of inequalities for experimentally accessible correlation functions (for a detailed derivation see Appendix II). In particular, if we let Alice choose her observable from the set of two settings $\vec{a}_1$ and $\vec{a}_2$, and Bob from the set of three settings $\vec{b}_1$, $\vec{b}_2$ and $\vec{b}_3=\vec{a}_2$, the following generalized Leggett-type inequality is obtained:
\begin{equation}
S_{NLHV}=|E_{11}(\varphi)+E_{23}(0)|+|E_{22}(\varphi)+E_{23}(0)|\leq4-\frac{4}{\pi}|\sin\frac{\varphi}{2}|\label{NLHV}
\end{equation}
where $E_{kl}(\varphi)$ is a uniform average of all correlation functions, defined in the plane of $\vec{a}_k$ and $\vec{b}_l$, with the same relative angle $\varphi$; the subscript NLHV stands for 'non-local hidden-variables'. For the inequality to be applied, vectors $\vec{a}_1$ and $\vec{b}_1$ necessarily have to lie in a plane orthogonal to the one defined by $\vec{a}_2$ and $\vec{b}_2$. This contrasts with the standard experimental configuration used to test the CHSH inequality, which is maximally violated for settings in one plane.\\

\begin{figure}[htbp]
\centerline{\includegraphics[width=0.8\textwidth]{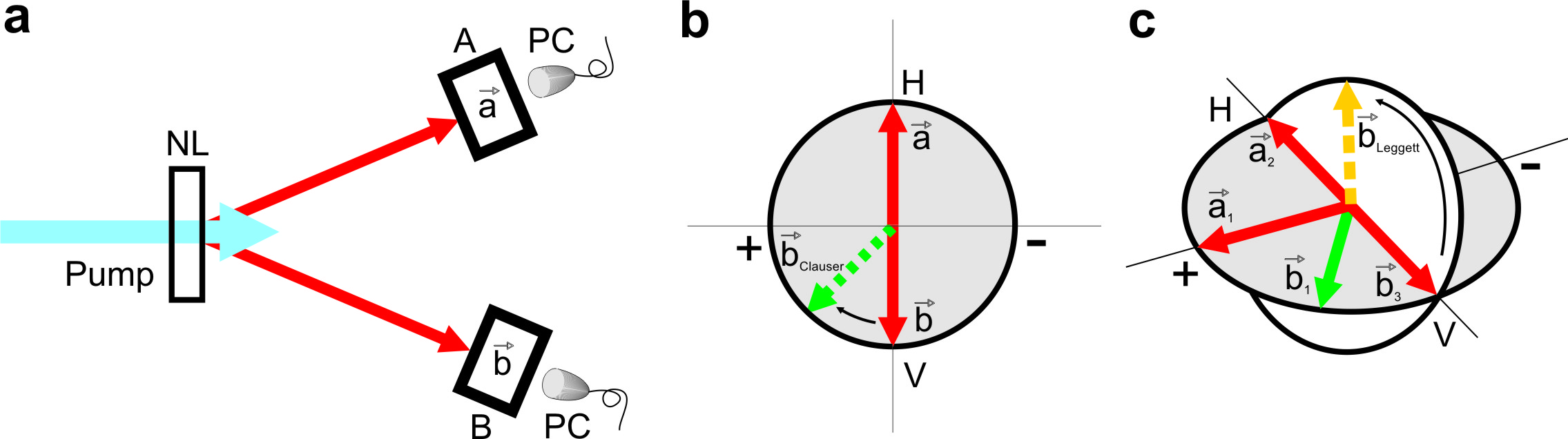}} \caption{\textbf{Testing non-local hidden-variable theories.} \textbf{a}, Diagram of a standard two-photon experiment to test for hidden-variable theories. When pumping a nonlinear crystal (NL) with a strong pump field, photon pairs are
created via SPDC and their polarization is detected with single-photon counters (PC). Local measurements at A and B are performed along directions $\vec{a}$ and $\vec{b}$ on the Poincar\'{e} sphere, respectively. Depending on the measurement directions, the obtained correlations can be used to test Bell inequalities (\textbf{b}) or Leggett-type inequalities (\textbf{c}). \textbf{b}, Correlations in one plane. Shown are measurements along directions in the linear plane of the Poincar\'{e} sphere (H (V) denotes horizontal (vertical) polarization). The original experiments by Wu and Shaknov~\cite{Wu1950} and Kocher and Commins~\cite{Kocher1967}, designed to test quantum predictions for correlated photon pairs, measured perfect correlations (solid lines). Measurements along the dashed line allow a Bell test, as was first performed by Freedman and Clauser~\cite{Freedman1972}. \textbf{c}, Correlations in orthogonal planes. All current experimental tests to violate Bell's inequality (CHSH) are performed within the shaded plane. Out-of-plane measurements are required for a direct test of the class of non-local hidden-variable theories, as was first suggested by Leggett.} \label{Figure1}
\end{figure}

The situation resembles in a way the status of the Einstein, Podolsky and Rosen (EPR) paradox before the advent of Bell's theorem and its first experimental tests. The experiments of Wu and Shaknov~\cite{Wu1950} and of Kocher and Commins~\cite{Kocher1967} were designed to demonstrate the validity of a quantum description of photon-pair correlations. As this task only required the testing of correlations along the same polarization direction, their results could not provide experimental data for the newly derived Bell inequalities (Fig.~\ref{Figure1}a, b). Curiously, as was shown by Clauser, Horne, Shimony and Holt, only a small modification of the measurement directions, such that non-perfect correlations of an entangled state are probed, was sufficient to test Bell's inequalities. The seminal experiment by Freedman and Clauser~\cite{Freedman1972} was the first direct and successful test~\cite{Clauser1978}. Today, all Bell tests - that is, tests of local realism - are performed by testing correlations of measurements along directions that lie in the same plane of the Poincar\'{e} sphere. Similar to the previous case, violation of the Leggett-type inequality requires only small modifications to that arrangement:\ To test the inequality, correlations of measurements along two orthogonal planes have to be probed (Fig.~\ref{Figure1}c). Therefore the existing data of all Bell tests cannot be used to test the class of nonlocal theories considered here.\\

Quantum theory violates inequality (\ref{NLHV}). Consider the quantum predictions for the polarization singlet state of two photons, $|\Psi^{-}\rangle_{AB}=\frac{1}{\sqrt{2}}\left[|H\rangle_A|V\rangle_B-|V\rangle_A|H\rangle_B\right]$, where, for example, $|H\rangle_A$ denotes a horizontally polarized photon propagating to Alice. The quantum correlation function for the measurements $\vec{a}_k$ and $\vec{b}_l$ performed on photons depends only on the relative angles between these vectors, and therefore $E_{kl}=-\vec{a}_k\cdot\vec{b}_l=-\cos\varphi$. Thus the left hand side of inequality (\ref{NLHV}), for quantum predictions, reads $|2(\cos\varphi+1)|$. The maximal violation of inequality (\ref{NLHV}) is for $\varphi_{max}=18.8^\circ$. For this difference angle, the bound given by inequality (\ref{NLHV}) equals 3.792 and the quantum value is 3.893.\\

Although this excludes the non-local models, it might still be possible that the obtained correlations could be explained by a local realistic model. In order to avoid that, we have to exclude both local realistic and non-local realistic hidden-variable theories. Note however that such local realistic theories need not be constrained by assumptions (1)-(3). The violation of the CHSH inequality invalidates all local realistic models. If one takes
\begin{equation}
S_{CHSH}=|E_{11}+E_{12}-E_{21}+E_{22}|\leq2 \label{CHSH}
\end{equation}
the quantum value of the left hand side for the settings used to maximally violate inequality (\ref{NLHV}) is 2.2156.\\

The correlation function determined in an actual experiment is typically reduced by a visibility factor $V$ to $E^{exp}=-V\cos\varphi$ owing to noise and imperfections. Thus to observe violations of inequality (\ref{NLHV}) (and inequality (\ref{CHSH})) in the experiment, one must have a sufficiently high experimental visibility of the observed interference. For the optimal difference angle $\varphi_{max}=18.8^{\circ}$, the minimum required visibility is given by the ratio of the bound (3.792) and the quantum value (3.893) of inequality (\ref{NLHV}), or $\sim97.4\%$. We note that in standard Bell-type experiments, a minimum visibility of only $\sim71\%$ is sufficient to violate the CHSH inequality, inequality (\ref{CHSH}), at the optimal settings. For the settings used here, the critical visibility reads $2/2.2156\approx90.3\%$, which is much lower than $97.4\%$.\\

\begin{figure}[htbp]
\centerline{\includegraphics[width=0.8\textwidth]{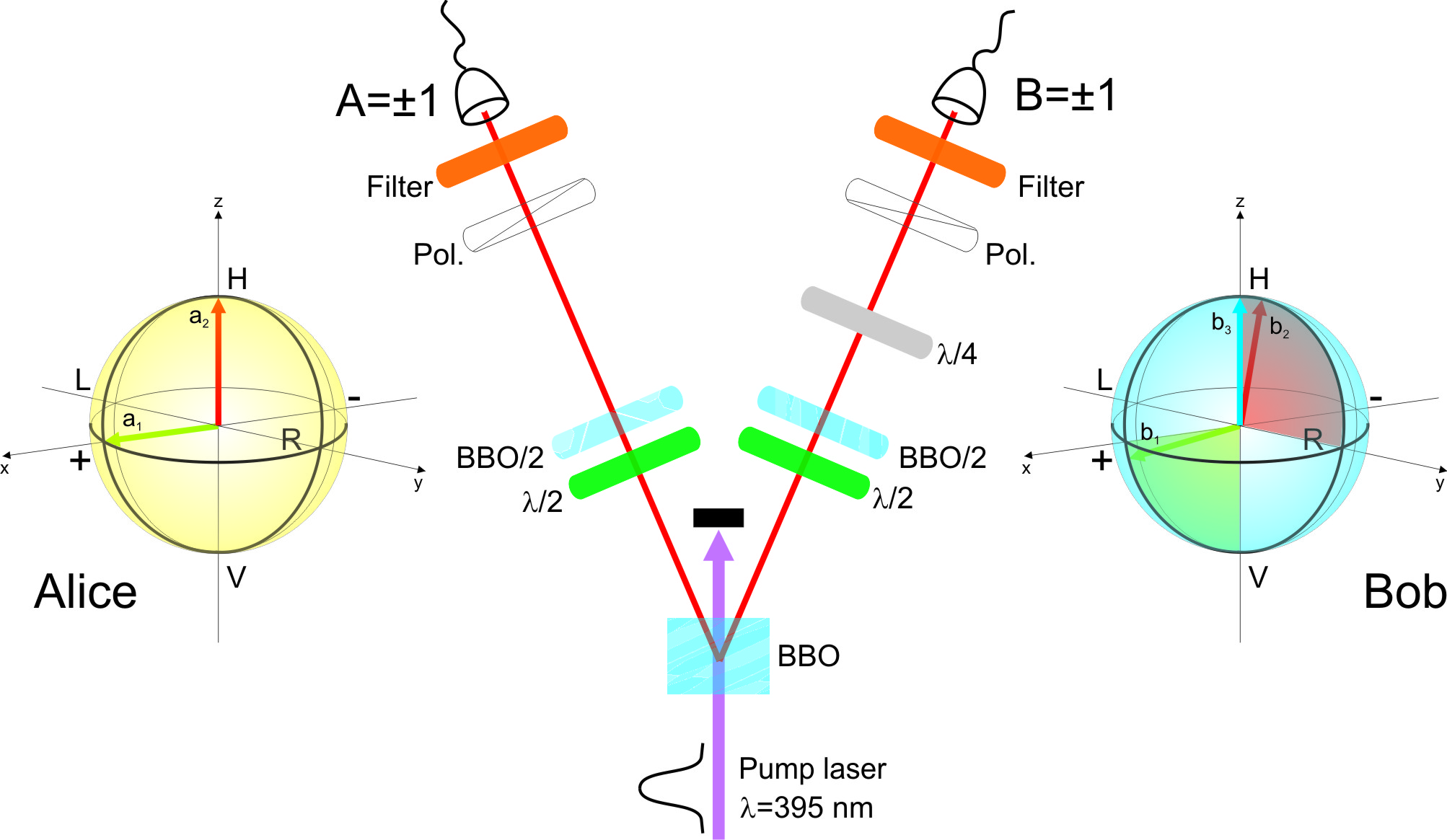}} \caption{\textbf{Experimental set-up.} A 2-mm-thick type-II $\beta$-barium-borate
(BBO) crystal is pumped with a pulsed frequency-doubled Ti:sapphire laser (180 fs) at $\lambda=395$~nm wavelength and $~\sim150$~mW optical c.w.\ power. The crystal is aligned to produce the polarization-entangled singlet state $|\Psi^{-}\rangle_{AB}=\frac{1}{\sqrt{2}}\left[|H\rangle_A|V\rangle_B-|V\rangle_A|H\rangle_B\right]$. Spatial and temporal distinguishability of the produced photons (induced by birefringence in the BBO) are compensated by a combination of half-wave plates ($\lambda$/2) and additional BBO crystals (BBO/2), while spectral distinguishability (due to the broad spectrum of the pulsed pump) is eliminated by narrow spectral filtering of $1$~nm bandwidth in front of each detector. In addition, the reduced pump power diminishes higher-order SPDC emissions of multiple photon pairs. This allows us to achieve a two-photon visibility of about $99\%$, which is well beyond the required threshold of $97.4\%$. The arrows in the Poincar\'{e} spheres indicate the measurement settings of Alice's and Bob's polarizers for the maximal violation of inequality (\ref{NLHV}). Note that setting $\vec{b}_2$ lies in the $y$-$z$ plane and therefore a quarter-wave plate has to be introduced on Bob's side. The coloured planes indicate the measurement directions for various difference angles $\varphi$ for both inequalities.} \label{Figure2}
\end{figure}

In the experiment (see Fig.~\ref{Figure2}), we generate pairs of polarization entangled photons via spontaneous parametric down-conversion (SPDC). The photon source is aligned to produce pairs in the polarization singlet state. We observed maximal coincidence count rates (per 10~s), in the $H/V$ basis, of around 3,500 with single count rates of 95,000 (Alice) and 105,000 (Bob), 3,300 coincidences in the $\pm45^{\circ}$ basis (75,000 singles at Alice and 90,000 at Bob), and 2,400 coincidences in the $R/L$ basis (70,000 singles at Alice and 70,000 at Bob). The reduced count rates in the $R/L$ basis are due to additional retarding elements in the beam path. The two-photon visibilities are approximately $99.0\pm1.2\%$ in the $H/V$ basis, $99.2\pm1.6\%$ in the $\pm45^{\circ}$ basis and $98.9\pm1.7\%$ in the $R/L$ basis, which - to our knowledge - is the highest reported visibility for a pulsed SPDC scheme. So far, no experimental evidence against the rotational invariance of the singlet state exists. We therefore replace the rotation averaged correlation functions in inequality (\ref{NLHV}) with their values measured for one pair of settings (in the given plane).\\

In terms of experimental count rates, the correlation function $E(\vec{a},\vec{b})$ for a given pair of general measurement settings is defined by
\begin{equation}
E(\vec{a},\vec{b})=\frac{N_{++}+N_{--}-N_{+-}-N_{-+}}{N_{++}+N_{--}+N_{+-}+N_{-+}}
\end{equation}
where $N_{AB}$ denotes the number of coincident detection events between Alice's and Bob's measurements within the integration time. We ascribe the number $+1$, if Alice (Bob) detects a photon polarized along $\vec{a}$ ($\vec{b}$), and $-1$ for the orthogonal direction $\vec{a}^{\perp}$ ($\vec{b}^{\perp}$). For example, $N_{+-}$ denotes the number of coincidences in which Alice obtains $\vec{a}$ and Bob $\vec{b}^{\perp}$. Note that $E(\vec{a}_k ,\vec{b}_l)=E_{kl}(\varphi)$, where $\varphi$ is the difference angle between the vectors $\vec{a}$ and $\vec{b}$ on the Poincar\'{e} sphere.\\

To test inequality (\ref{NLHV}), three correlation functions ($E(\vec{a}_1 ,\vec{b}_1,)$, $E(\vec{a}_2 ,\vec{b}_2)$, $E(\vec{a}_2 ,\vec{b}_3)$) have to be extracted from the measured data. We choose observables $\vec{a}_1$ and $\vec{b}_1$ as linear polarization measurements (in the $x$-$z$ plane on the Poincar\'{e} sphere; see Fig.~\ref{Figure2}) and $\vec{a}_2$ and $\vec{b}_2$ as elliptical polarization measurements in the $y$-$z$ plane. Two further correlation functions ($E(\vec{a}_2,\vec{b}_1)$ and $E(\vec{a}_1,\vec{b}_2)$) are extracted to test the CHSH inequality, inequality (\ref{CHSH}).\\

\begin{figure}[htbp]
\centerline{\includegraphics[width=0.8\textwidth]{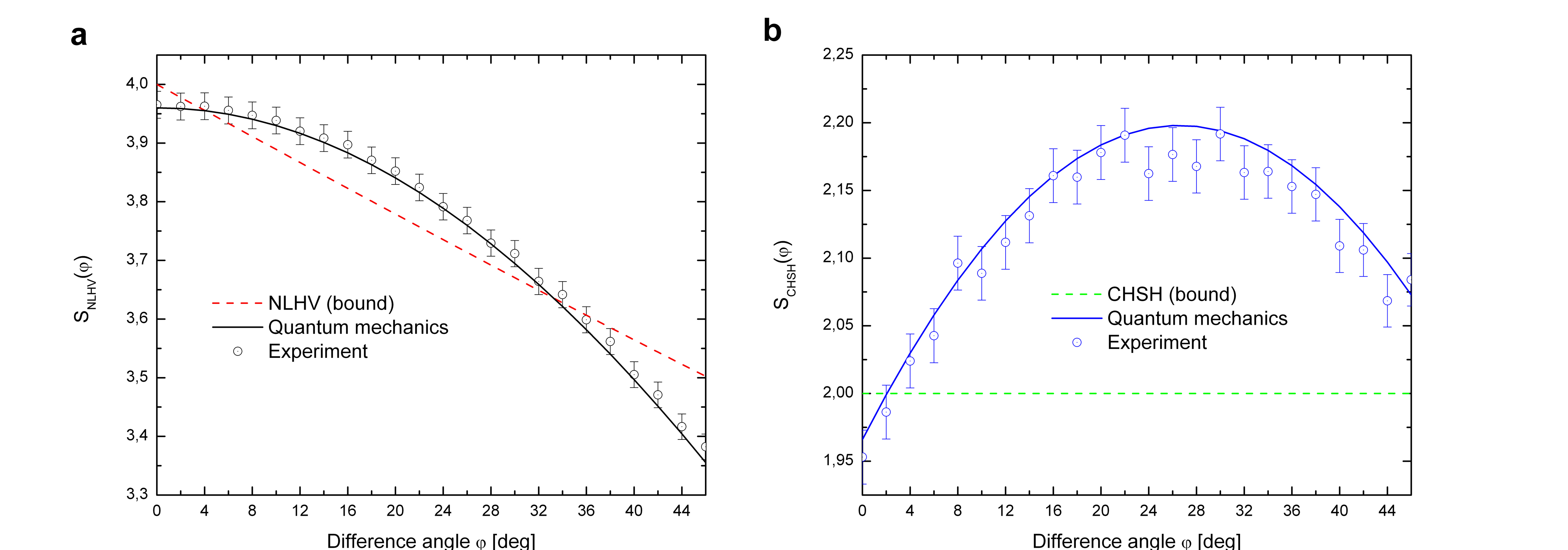}} \caption{\textbf{Experimental violation of the inequalities for non-local hidden-variable theories (NLHV) and for local realistic theories (CHSH).} \textbf{a}, Dashed line indicates the bound of inequality (\ref{NLHV}) for the investigated class of nonlocal hidden-variable theories (see text). The solid line is the quantum theoretical prediction reduced by the experimental visibility. The shown experimental data were taken for various difference angles $\varphi$ (on the Poincar\'{e} sphere) of local measurement settings. The bound is clearly violated for $4^{\circ}<\varphi<36^{\circ}$. Maximum violation is observed for $\varphi_{max}\approx20^{\circ}$. \textbf{b}, At the same time, no local realistic theory can model the correlations for the investigated settings as the same set of data also violates the CHSH inequality (\ref{CHSH}). The bound (dashed line) is overcome for all values $\varphi$ around $\varphi_{max}$, and hence excludes any local realistic explanation of the observed correlations in \textbf{a}. Again, the solid line is the quantum prediction for the observed experimental visibility. Error bars indicate s.d.} \label{Figure3}
\end{figure}

The first set of correlations, in the $x$-$z$ plane, is obtained by using linear polarizers set to $\alpha_1$ and $\beta_1$ (relative to the $z$-axis) at Alice's and Bob's location, respectively. In particular, $\alpha_1=\pm45^{\circ}$, while $\beta_1$ is chosen to lie between $45^{\circ}$ and $160^{\circ}$ (green arrows in Fig.~\ref{Figure2}). The second set of correlations (necessary for CHSH) is obtained in the same plane for $\alpha_2=0^{\circ}/90^{\circ}$ and $\beta_1$ between $45^{\circ}$ and $160^{\circ}$. The set of correlations for measurements in the $y$-$z$ plane is obtained by introducing a quarter-wave plate with the fast axis aligned along the (horizontal) $0^{\circ}$-direction at Bob's site, which effectively rotates the polarization state by $90^{\circ}$ around the $z$-axis on the Poincar\'{e} sphere (red arrows in Fig.~\ref{Figure2}). The polarizer angles are then set to $\alpha_2=0^{\circ}/90^{\circ}$ and $\beta_2$ is scanned between $0^{\circ}$ and $115^{\circ}$. With the same $\beta_2$ and $\alpha_1=\pm45^{\circ}$, the expectation values specific only for the CHSH case are measured. The remaining measurement for inequality (\ref{NLHV}) is the check of perfect correlations, for which we choose $\alpha_2=\beta_3=0^{\circ}$, that is, the intersection of the two orthogonal planes. Figure~\ref{Figure3} shows the experimental violation of inequalities (\ref{NLHV}) and (\ref{CHSH}) for various difference angles. Maximum violation of inequality (\ref{NLHV}) is achieved, for example, for the settings $\{\alpha_1,\alpha_2,\beta_1,\beta_2,\beta_3\}=\{45^{\circ},0^{\circ},55^{\circ},10^{\circ},0^{\circ}\}$.\\

We finally obtain the following expectation values for a difference angle $\varphi=20^{\circ}$ (the errors are calculated assuming that the counts follow a poissonian distribution):\ $E(\vec{a}_1,\vec{b}_1)=-0.9298\pm0.0105$, $E(\vec{a}_2,\vec{b}_2)=-0.942\pm0.0112$, $E(\vec{a}_2,\vec{b}_3)=-0.9902\pm0.0118$. This results in $S_{NLHV}=3.8521\pm0.0227$, which violates inequality (\ref{NLHV}) by $3.2$ standard deviations (see Fig.~\ref{Figure3}). At the same time, we can extract the additional correlation functions $E(\vec{a}_2,\vec{b}_1)=0.3436\pm0.0088$, $E(\vec{a}_1,\vec{b}_2)=0.0374\pm0.0091$ required for the CHSH inequality. We obtain $S_{CHSH}=2.178\pm0.0199$, which is a violation by $\sim9$ standard deviations. The stronger violation of inequality (\ref{CHSH}) is due to the relaxed visibility requirements on the probed entangled state.\\

We have experimentally excluded a class of important non-local hidden-variable theories. In an attempt to model quantum correlations of entangled states, the theories under consideration assume realism, a source emitting classical mixtures of polarized particles (for which Malus' law is valid) and arbitrary non-local dependencies via the measurement devices. Besides their natural assumptions, the main appealing feature of these theories is that they allow us both to model perfect correlations of entangled states and to explain all existing Bell-type experiments. We believe that the experimental exclusion of this particular class indicates that any non-local extension of quantum theory has to be highly counterintuitive. For example, the concept of ensembles of particles carrying definite polarization could fail. Furthermore, one could consider the breakdown of other assumptions that are implicit in our reasoning leading to the inequality. These include Aristotelian logic, counterfactual definiteness, absence of actions into the past or a world that is not completely deterministic~\cite{Bell1985}. We believe that our results lend strong support to the view that any future extension of quantum theory that is in agreement with experiments must abandon certain features of realistic descriptions.\\
\\

\begin{acknowledgements}
We are grateful to A. J. Leggett for stimulating this work and for discussions. We also thank D. Greenberger, M. A. Horne, T. Jennewein, J. Kofler, S. Malin and A. Shimony for their comments. T.P. is grateful for the hospitality of the IQOQI, Vienna. M.A. thanks L. Gohlike for his hospitality at the Seven Pines VIII. We acknowledge support from the Austrian Science Fund (FWF), the European Commission, the Austrian Exchange Service (\"OAD), the Foundation for Polish Science (FNP), the Polish Ministry of Higher Education and Science, the City of Vienna and the Foundational Questions Institute (FQXi).
\end{acknowledgements}

\newpage
\appendix
\section{Appendix I: An explicit non-local hidden-variable model}

We construct an explicit non-local model compliant with the
introduced assumptions (1)-(3). It perfectly simulates all quantum
mechanical predictions for measurements in a plane of the Poincar\'e
sphere. We model the correlation function of the singlet state,
$E_{\vec a \vec b}^{QM} = -\vec a \cdot \vec b$, for which all local
averages $\langle A \rangle$ and $\langle B \rangle$ vanish. In
particular, the violation of any CHSH-type inequality can be
explained within the model and, in addition, all perfect
correlations state can be recovered.

Let us start with a source that emits photons with well-defined
polarization. Polarization $\vec u$ is sent to Alice and $\vec v$ to
Bob. Alice sets her measuring device to $\vec a$ and Bob to $\vec
b$. The hidden-variable $\lambda \in [0,1]$ is carried by both
particles and predetermines the individual measurement result as
follows:
\begin{eqnarray}
A \equiv A(\vec a, \vec u, \lambda) &=& \Big\{
\begin{array}{ccc}
+1 & \textrm{ for } & \lambda \in [0,\lambda_A], \\
-1 & \textrm{ for } & \lambda \in (\lambda_A,1],
\end{array}
\textrm{ with } \lambda_A = \frac{1}{2} (1 + \vec u \cdot \vec a),
\label{A_DEF}
\end{eqnarray}
where $A$ is the outcome of Alice. This means, whenever
$\lambda\leq\lambda_A$ the result of the measurement $A$ is $+1$,
and for $\lambda >\lambda_A$ the result is $-1$. Note that the
measurement settings only enter in $\lambda_A$ and are hence
independent of the hidden-variable $\lambda$ of the source. The
outcome of Bob is given by
\begin{eqnarray}
B \equiv B(\vec a, \vec b, \vec u, \vec v, \lambda) &=& \Big\{
\begin{array}{ccc}
+1 & \textrm{ for } & \lambda \in [x_1,x_2], \\
-1 & \textrm{ for } & \lambda \in [0,x_1) \cup (x_2,1],
\end{array}
\textrm{ with } x_1,x_2 \in [0,1] \textrm{ arbitrary but } x_2-x_1 =
\frac{1}{2}(1+\vec v \cdot \vec b).
\end{eqnarray}
All non-local dependencies are put on the side of Bob. His measuring
device has the information about the setting of Alice, $\vec a$, and
her polarization $\vec u$. The requirement of the non-local models
discussed here is that the local averages performed on the
subensemble of definite (but arbitrary) polarizations $\vec u$ and
$\vec v$ obey Malus' Law, i.e. $\overline{A_{\vec u}} = \vec u \cdot
\vec a$ for Alice, and $\overline{B_{\vec v}} = \vec v \cdot \vec b$
for Bob. Indeed, a straight-forward calculation shows that this
requirement is fulfilled for both Alice and Bob:
\begin{equation}
\overline{A_{\vec u}} = \int_0^{\lambda_A} d \lambda -
\int_{\lambda_A}^1 d \lambda = 2 \lambda_A - 1 = \vec u \cdot
\vec a,
\end{equation}
\begin{equation}
\overline{B_{\vec v}} = \int_{x_1}^{x_2} d \lambda - \int_{0}^{x_1} d \lambda - \int_{x_2}^{1} d \lambda
= 2 (x_2-x_1) - 1 = \vec v \cdot \vec b.
\label{BOB_MALUS}
\end{equation}
Thus the current construction fulfills one of our aims -- we recover
Malus' law. In order to get the correct formula for correlated
counts one can fix the value of $x_1$ and $x_2$ in the following
way:
\begin{eqnarray}
x_1 &=& \frac{1}{4}[1 + \vec u \cdot \vec a - \vec v \cdot \vec b + \vec a \cdot \vec b], \nonumber \\
x_2 &=& \frac{1}{4}[3 + \vec u \cdot \vec a + \vec v \cdot \vec b
+ \vec a \cdot \vec b]. \label{xes}
\end{eqnarray}
With this definitions whenever $x_1 \le \lambda_A \le x_2$ the
expectation value for measurements on the subensembles reproduces
quantum correlations. Simply:
\begin{equation}
\overline{A_{\vec u}B_{\vec v}} = - \int_0^{x_1} d \lambda +
\int_{x_1}^{\lambda_A} d \lambda - \int_{\lambda_A}^{x_2} d
\lambda + \int_{x_2}^{1} d \lambda = 2(\lambda_A - x_1 - x_2) +
1=-\vec a \cdot \vec b.
\end{equation}
Therefore, in the next step, one must find the conditions for which
both $x_1$ and $x_2$ take values from $[0,1]$ and $x_1$ and $x_2$
take values from $[0,1]$.

To this end, using definitions (\ref{xes}) one finds that the first
condition is equivalent to a set of four inequalities:
\begin{eqnarray}
-1 + \vec v \cdot \vec b \le & \vec a \cdot \vec b + \vec u \cdot \vec a & \le 3 + \vec v \cdot \vec b, \nonumber \\
-3 - \vec v \cdot \vec b \le & \vec a \cdot \vec b + \vec u \cdot \vec a & \le 1 - \vec v \cdot \vec b.
\end{eqnarray}
Note that the upper bound $3 + \vec v \cdot \vec b$
cannot be exceeded by the middle term,
as well as the lower bound $-3 - \vec v \cdot \vec b$.
Thus, this set of four inequalities is equivalent
to a single one:
\begin{equation}
|\vec a \cdot \vec b + \vec u \cdot \vec a|  \le 1 - \vec v \cdot \vec b.
\label{COMPLEMENTARITY1}
\end{equation}
Similarly, the second condition can be reexpressed as:
\begin{equation}
|\vec a \cdot \vec b - \vec u \cdot \vec a|  \le 1 + \vec v \cdot \vec b.
\label{COMPLEMENTARITY2}
\end{equation}
Finally, the validity condition for the model
is a conjunction of (\ref{COMPLEMENTARITY1}) and (\ref{COMPLEMENTARITY2}):
\begin{equation}
|\vec a \cdot \vec b \pm \vec u \cdot \vec a|  \le 1 \mp \vec v \cdot \vec b.
\label{COMPLEMENTARITY}
\end{equation}
If this relation  is not satisfied the model does not recover
quantum correlations. Either it becomes inconsistent since $x_1$ or
$x_2$ leave their range or the necessary relation $x_1 \le \lambda_A
\le x_2$ is not satisfied, or both. This is the origin of the
incompatibility with general quantum predictions. Nevertheless the
model can explain all perfect correlations and the violation of the
CHSH inequality.

Imagine a source producing pairs with the following property:
whenever polarization $\vec u$ is sent to Alice polarization $\vec v
= - \vec u$ is sent to Bob. Both parties locally observe random
polarizations. For Alice, the local average over different
polarizations yields
\begin{equation}
\langle A \rangle = \frac{1}{2} \overline{A_{\vec u}} +  \frac{1}{2}\overline{A_{-\vec
u}} = \frac{1}{2}\vec u \cdot \vec a - \frac{1}{2}\vec u \cdot \vec a = 0,
\end{equation}
as it should be for the singlet state. The same result holds for
Bob. In this way, we have reproduced the randomness of local
measurement outcomes, typical for measurements on entangled
states.

With the same source, one can explain perfect correlations for
measurements along the same basis, i.e. $\vec b = \pm \vec a$. To
see how the model works take $\vec v = - \vec u$ and $\vec b = \vec
a$, and find that $x_1 = \frac{1}{2}[1 + \vec u \cdot \vec a] =
\lambda_A$ and $x_2 = 1$. As it should be, Bob's outcomes are always
opposite to Alice's:
\begin{eqnarray}
B \equiv B(\vec a, \vec a, \vec u, - \vec u, \lambda) &=& \Big\{
\begin{array}{ccc}
+1 & \textrm{ for } & \lambda \in [\lambda_A,1], \\
-1 & \textrm{ for } & \lambda \in [0,\lambda_A).
\end{array}
\end{eqnarray}
If in the same subensemble we take $\vec b = - \vec a$ we obtain
$x_1 = 0$ and $x_2 = \lambda_A$, which results in $B=A$, again in
full agreement with quantum mechanics. Note that for these
measurement settings, i.e. in order to obtain perfect correlations,
condition (\ref{COMPLEMENTARITY}) imposes no additional restrictions
since it is always satisfied. For example, if $\vec u$ is sent to
Alice and $\vec b = -\vec a$ one obtains $|-1 \pm \vec u \cdot \vec
a| \le 1 \mp \vec u \cdot \vec a$, which always holds. The same
argument applies to the other subensemble and other measurement
possibilities $\vec b = \pm \vec a$.

Finally, the full predictions of quantum theory are recovered if
Alice and Bob restrict their measurements to lie in planes
orthogonal to vectors $\vec u$ and $\vec v$, respectively, i.e.
$\vec u \cdot \vec a = \vec v \cdot \vec b = 0$. In this case,
condition (\ref{COMPLEMENTARITY}) is satisfied, as $|\vec a \cdot
\vec b| \le 1$. In general, if condition (\ref{COMPLEMENTARITY}) is
satisfied, i.e. for a consistent set of parameters, our model
reproduces quantum correlations since they are already reproduced in
every subensemble and hence averaging over different polarizations
is not necessary:
\begin{equation}
\langle AB \rangle = \overline{A_{\vec u}B_{-\vec u}} =
\overline{A_{-\vec u}B_{\vec u}} = - \vec a \cdot \vec b.
\end{equation}
Therefore every experimental violation of the CHSH inequality can be
explained by the presented non-local model.

\section{Appendix II: Derivation of the inequality}

After Leggett~\cite{Leggett2003}, one can take a source which distributes pairs of well-polarized photons. Different pairs can have different polarizations. The size of a subensemble in which photons have polarizations $\vec u$ and $\vec v$ is described by the weight function $F(\vec u, \vec v)$. All polarizations and measurement directions are represented as vectors on the Poincar\'{e} sphere. In every such subensemble individual measurement outcomes are determined by hidden variables $\lambda$. The hidden variables are distributed according to the distribution $\rho_{\vec u, \vec v}(\lambda)$.

For any dichotomic measurement results, $A=\pm 1$ and $B=\pm 1$, the following identity holds:
\begin{equation}
-1 + |A+B| = A B = 1 - |A - B|. \label{NL_IDENTITY}
\end{equation}
If the signs of $A$ and $B$ are the same $|A+B| = 2$ and $|A-B|=0$, and if $A = -B$ then $|A+B| = 0$ and $|A-B|=2$.
Any kind of non-local dependencies is allowed, i.e. $A = A(\vec a, \vec b, \vec u, \vec v, \lambda, ....)$ and $B = B(\vec a, \vec b, \vec u, \vec v, \lambda, ....)$. Taking the average over the subensemble with definite polarizations gives:
\begin{equation}
-1 + \int d \lambda \rho_{\vec u, \vec v}(\lambda) |A+B| = \int d
\lambda \rho_{\vec u, \vec v}(\lambda) AB = 1 - \int d \lambda
\rho_{\vec u, \vec v}(\lambda)|A-B|,
\end{equation}
which in an abbreviated notation, where the averages are denoted by bars, is
\begin{equation}
-1 + \overline{|A + B|} = \overline{AB} = 1 - \overline{|A - B|}.
\end{equation}
Since the average of the modulus is greater or equal to the modulus of the averages one gets the set of inequalities
\begin{equation}
-1 + |\overline A + \overline B| \le \overline{AB} \le 1 -
|\overline A - \overline B|. \label{SET_OF_INEQ}
\end{equation}
From now on only the upper bound will be considered, however all steps apply to the lower bound as well.

With the assumption that photons with well defined polarization obey Malus' law:
\begin{equation}
\overline A = \vec u \cdot \vec a, \qquad
\overline B = \vec v \cdot \vec b, \label{LOCAL_AVERAGES}
\end{equation}
the upper bound of Eq.~(\ref{SET_OF_INEQ}) becomes:
\begin{equation}
\overline{AB} \le 1 - |\vec u \cdot \vec a_k - \vec v \cdot \vec
b_l|,
\end{equation}
where $\vec a_k$ and $\vec b_l$ are unit vectors associated with the $k$th measurement setting of Alice and the $l$th of Bob, respectively.

Taking the average over arbitrary polarizations one obtains
\begin{equation}
E_{kl} \le 1 - \int d \vec u d \vec v F(\vec u, \vec v)
 |\vec u \cdot \vec a_k - \vec v \cdot \vec b_l|,
\end{equation}
where $E_{kl}$ is the correlation function which can be experimentally measured when Alice chooses to measure $\vec a_k$ and Bob chooses $\vec b_l$. Let us denote by $u_{kl}$ and $v_{kl}$ the length of projections of vectors $\vec u$
and $\vec v$ onto the plane spanned by $\vec a_k$ and $\vec b_l$. Since one can decompose vectors $\vec u$ and $\vec v$ into a vector orthogonal to the plane of the settings and a vector within the plane the scalar products read:
\begin{eqnarray}
\vec u \cdot \vec a_k &=& u_{kl} \cos(\phi_{a_k} - \phi_u), \\
\vec v \cdot \vec b_l &=& v_{kl} \cos(\phi_{b_l} - \phi_v),
\end{eqnarray}
where all the $\phi$ angles are relative to some axis within the plane of the settings; Angles $\phi_u$ and $\phi_v$ describe the position of the projections of vectors $\vec u$ and $\vec v$, respectively, whereas angles $\phi_{a_k}$ and $\phi_{b_l}$ describe the position of the setting vectors. With this notation the inequality transforms to:
\begin{eqnarray}
E_{kl} & \le & 1 - \int d \vec u d \vec v F(\vec u, \vec v)
|u_{kl} \cos(\phi_{a_k} - \phi_u) - v_{kl} \cos(\phi_{b_l} - \phi_v)|.
\end{eqnarray}
The magnitudes of the projections can always be decomposed into the sum and the difference of two real numbers:
\begin{equation}
u_{kl} = n_1 + n_2, \qquad v_{kl} = n_1 - n_2.
\end{equation}
We insert this decomposition into the last inequality, and hence the terms multiplied by $n_1$ and $n_2$ read:
\begin{eqnarray}
\cos(\phi_{a_k} - \phi_u) & - & \cos(\phi_{b_l} - \phi_v) \nonumber \\
&=& 2 \sin \frac{\phi_{a_k} + \phi_{b_l} - (\phi_u + \phi_v)}{2}
\sin\frac{-(\phi_{a_k} - \phi_{b_l})+\phi_u - \phi_v}{2}, \\
\cos(\phi_{a_k} - \phi_u) &+& \cos(\phi_{b_l} - \phi_v)  \nonumber \\
&=& 2 \cos \frac{\phi_{a_k} + \phi_{b_l} - (\phi_u + \phi_v)}{2}
\cos \frac{\phi_{a_k} - \phi_{b_l} - (\phi_u - \phi_v)}{2},
\end{eqnarray}
respectively. We make the following substitution for the measurement angles:
\begin{eqnarray}
\xi_{kl} = \frac{\phi_{a_k} + \phi_{b_l}}{2}, \qquad
&& \varphi_{kl} = \phi_{a_k} - \phi_{b_l},
\end{eqnarray}
and parameterize the position of the projections within their plane by:
\begin{eqnarray}
\psi_{uv} = \frac{\phi_u + \phi_v}{2}, \qquad
&& \chi_{uv} = \phi_u - \phi_v.
\end{eqnarray}
Using these new angles one arrives at:
\begin{eqnarray}
E_{kl}(\xi_{kl},\varphi_{kl}) \le 1- 2 \int d \vec u d \vec v F(\vec u, \vec v)
|n_2 \cos\frac{\varphi_{kl} - \chi_{uv}}{2} \cos (\xi_{kl}-\psi_{uv}) - n_1 \sin
\frac{\varphi_{kl} - \chi_{uv}}{2} \sin (\xi_{kl}-\psi_{uv})|,
\end{eqnarray}
where in the correlation function $E_{kl}(\xi_{kl},\varphi_{kl})$ we explicitly state the angles it is dependent on. The expression within the modulus is a linear combination of two harmonic functions of $\xi_{kl} - \psi_{uv}$, and therefore is a harmonic function itself. Its amplitude reads $\sqrt{n_2^2 \cos^2(\frac{\varphi_{kl} - \chi_{uv}}{2})+ n_1^2 \sin^2(\frac{\varphi_{kl} - \chi_{uv}}{2})}$, and the phase is some fixed real number $\alpha$:
\begin{eqnarray}
E_{kl}(\xi_{kl}, \varphi_{kl}) \le 1 - 2
\int d \vec u d \vec v F(\vec u, \vec v)
\sqrt{n_2^2
\cos^2(\frac{\varphi_{kl} - \chi_{uv}}{2}) + n_1^2 \sin^2(\frac{\varphi_{kl} -
\chi_{uv}}{2})} |\cos(\xi_{kl} - \psi_{uv} + \alpha)|.
\label{XI_PHI_CORR}
\end{eqnarray}
In the next step we average both sides of this inequality over the \emph{measurement angle} $\xi_{kl} = \frac{\phi_{a_k} + \phi_{b_l}}{2}$. This means an integration over $\xi_{kl} \in [0, 2 \pi )$ and a multiplication by $\frac{1}{2 \pi}$. The integral of the $\xi_{kl}$ dependent part of the right-hand side of (\ref{XI_PHI_CORR}) reads:
\begin{equation}
\frac{1}{2 \pi} \int_0^{2 \pi} d \xi_{kl} |\cos(\xi_{kl} - \psi_{uv} + \alpha)| = \frac{2}{\pi}.
\end{equation}
By denoting the average of the correlation function over the angle $\xi_{kl}$ as:
\begin{equation}
E_{kl}(\varphi_{kl}) \equiv \frac{1}{2\pi} \int_0^{2 \pi} d \xi_{kl} E_{kl}(\xi_{kl}, \varphi_{kl}),
\end{equation}
one can write (\ref{XI_PHI_CORR}) as
\begin{equation}
E_{kl}(\varphi_{kl})  \le 1-\frac{4}{\pi}
\int d \vec u d \vec v F(\vec u, \vec v)
\sqrt{n_2^2 \cos^2(\frac{\varphi_{kl} - \chi_{uv}}{2}) + n_1^2 \sin^2(\frac{\varphi_{kl} - \chi_{uv}}{2})}.
\end{equation}
This inequality is valid for any choice of observables in the plane defined by $\vec a_k$ and $\vec b_l$. One can introduce two new observable vectors in this plane and write the inequality for the averaged correlation function $E_{k'l'}(\varphi_{k'l'}')$ of these new observables. The sum of these two inequalities is
\begin{eqnarray}
&&E_{kl}(\varphi_{kl}) + E_{k'l'}(\varphi_{k'l'}')  \le 2-\frac{4}{\pi}
\int d \vec u d \vec v F(\vec u, \vec v) \nonumber \\
&\times&  \left(
\sqrt{n_2^2 \cos^2(\frac{\varphi_{kl} - \chi_{uv}}{2}) + n_1^2 \sin^2(\frac{\varphi_{kl} - \chi_{uv}}{2})}
+ \sqrt{n_2^2 \cos^2(\frac{\varphi_{k'l'}' - \chi_{uv}}{2}) + n_1^2 \sin^2(\frac{\varphi_{k'l'}' - \chi_{uv}}{2})}
\right)
\label{21}
\end{eqnarray}
One can use the triangle inequality
\begin{eqnarray}\label{TRIANGLE_TRICK}
||\vec x + \vec y|| & \le & ||\vec x || + || \vec y ||, \\
\sqrt{(x_1+y_1)^2 + (x_2 + y_2)^2} & \le & \sqrt{x_1^2 + x_2^2} + \sqrt{y_1^2 + y_2^2},
\end{eqnarray}
for the two-dimensional vectors $\vec x = (x_1,x_2)$ and $\vec y = (y_1,y_2)$, with components defined by:
\begin{eqnarray}
x_1 = |n_2 \cos\frac{\varphi_{kl} - \chi_{uv}}{2}|, & \quad &
y_1 = |n_2 \cos\frac{\varphi_{k'l'}' - \chi_{uv}}{2}|, \\
x_2 = |n_1 \sin\frac{\varphi_{kl} - \chi_{uv}}{2}|, & \quad &
y_2 = |n_1 \sin\frac{\varphi_{k'l'}' - \chi_{uv}}{2}|.
\end{eqnarray}
This implies that the integrand is bounded from below by:
\begin{eqnarray}
\sqrt{n_2^2 \cos^2 \frac{\varphi_{kl} - \chi_{uv}}{2} + n_1^2
\sin^2 \frac{\varphi_{kl} - \chi_{uv}}{2}} +\sqrt{n_2^2 \cos^2
\frac{\varphi_{k'l'}' - \chi_{uv}}{2} + n_1^2 \sin^2 \frac{\varphi_{k'l'}' -
\chi_{uv}}{2}} \nonumber \\
\ge
\sqrt{n_2^2 \Big( |\cos\frac{\varphi_{kl} - \chi_{uv}}{2}| + |\cos\frac{\varphi_{k'l'}' - \chi_{uv}}{2}| \Big)^2
+ n_1^2 \Big( |\sin\frac{\varphi_{kl} - \chi_{uv}}{2}| + |\sin \frac{\varphi_{k'l'}' - \chi_{uv}}{2}| \Big)^2 }.
\end{eqnarray}
One can further estimate this bound by using the following relations
\begin{eqnarray}\label{NL_APPROX_BOUND2}
|\cos(\frac{\varphi_{kl} - \chi_{uv}}{2})| + |\cos(\frac{\varphi_{k'l'}' - \chi_{uv}}{2})| & \ge & |\sin\frac{\varphi_{kl}-\varphi_{k'l'}'}{2}|\;\;\textrm{and} \\
|\sin(\frac{\varphi_{kl} - \chi_{uv}}{2})| + |\sin(\frac{\varphi_{k'l'}' - \chi_{uv}}{2})| & \ge & |\sin\frac{\varphi_{kl}-\varphi_{k'l'}'}{2}|.
\label{NL_APPROX_BOUND}
\end{eqnarray}
This estimate follows if one uses the formula for the sine of the difference angle to the right-hand side argument $\frac{\varphi_{kl}-\varphi_{k'l'}'}{2} = \frac{\varphi_{kl}- \chi_{uv}}{2} - \frac{\varphi_{k'l'}' - \chi_{uv}}{2}$. Namely,
\begin{eqnarray}
|\sin\frac{\varphi_{kl}-\varphi_{k'l'}'}{2}| & = &
|\sin\frac{\varphi_{kl} - \chi_{uv}}{2} \cos\frac{\varphi_{k'l'}' - \chi_{uv}}{2} -
\cos\frac{\varphi_{kl} - \chi_{uv}}{2} \sin\frac{\varphi_{k'l'}' - \chi_{uv}}{2}| \\
& \le & |\sin\frac{\varphi_{kl} - \chi_{uv}}{2}| |\cos\frac{\varphi_{k'l'}' - \chi_{uv}}{2}| +
|\cos\frac{\varphi_{kl} - \chi_{uv}}{2}| |\sin\frac{\varphi_{k'l'}' - \chi_{uv}}{2}|.
\end{eqnarray}
After these estimates, the lower bound of $E_{kl}+E_{k'l'}$ (following form the left-hand side inequality in (\ref{SET_OF_INEQ})) is equal to minus the upper bound, and thus one can can apply the upper bound to the modulus of the left hand side of (\ref{21}). This is because the only formal difference between expressions in the estimates seeking the lower bound of the averaged Eq.~(\ref{SET_OF_INEQ}) compared to those seeking the upper bound boils down to the interchange between $n_1$ and $n_2$. After applying (\ref{NL_APPROX_BOUND2}) and~(\ref{NL_APPROX_BOUND}), this makes no difference anymore. One can shortly write:
\begin{eqnarray}
| E_{kl}(\varphi_{kl}) + E_{k'l'}(\varphi_{k'l'}') | & \le &  2
- \frac{4}{\pi} |\sin\frac{\varphi_{kl} - \varphi_{k'l'}'}{2}|  \int d \vec u d \vec v F(\vec u, \vec v) \sqrt{n_2^2 + n_1^2} .
\end{eqnarray}
Going back to the magnitudes:
\begin{eqnarray}
| E_{kl}(\varphi_{kl}) + E_{k'l'}(\varphi_{k'l'}') | & \le &  2
- \frac{2 \sqrt{2}}{\pi} |\sin\frac{\varphi_{kl} - \varphi_{k'l'}'}{2}|  \int d \vec u d \vec v F(\vec u, \vec v) \sqrt{u_{kl}^2 + v_{kl}^2} .
\label{XY}
\end{eqnarray}
This inequality is valid for \emph{any} choice of the plane of observables. The bound involves only the projections of vectors $\vec u$ and $\vec v$ onto the plane of the settings. The integrations in the bound can be thought of as a mean value of expression $\sqrt{u_{kl}^2 + v_{kl}^2}$ averaged over the distribution of the vectors. For the plane orthogonal to the initial one the inequality is
\begin{eqnarray}
| E_{pq}^{\perp}(\varphi_{pq}) + E_{p'q'}^{\perp}(\varphi_{p'q'}') |
& \le &  2
- \frac{2 \sqrt{2}}{\pi} |\sin\frac{\varphi_{pq} - \varphi_{p'q'}'}{2}|  \int d \vec u d \vec v F(\vec u, \vec v) \sqrt{u_{pq}^2 + v_{pq}^2} .
\label{XZ}
\end{eqnarray}
where $u_{pq}$ and $v_{pq}$ denote the projections of vectors $\vec u$ and $\vec v$, respectively, onto the plane spanned by the settings $\vec a_p$ and $\vec b_q$ (which is by construction orthogonal to the plane spanned by $\vec a_k$ and $\vec b_l$). We add the inequalities for orthogonal observation planes, (\ref{XY}) and (\ref{XZ}), and choose $\varphi_{k'l'}' = \varphi_{p'q'}' = 0$ and $\varphi_{kl} = \varphi_{pq} = \varphi$. This gives
\begin{equation}
| E_{kl}(\varphi) + E_{k'l'}(0) |
+ | E_{pq}^{\perp}(\varphi) + E_{p'q'}^{\perp}(0)  |
\le  4
- \frac{2 \sqrt{2}}{\pi} |\sin\frac{\varphi}{2}|  \int d \vec u d \vec v F(\vec u, \vec v)
\left( \sqrt{u_{kl}^2 + v_{kl}^2} + \sqrt{u_{pq}^2 + v_{pq}^2} \right).
\end{equation}
We apply the triangle inequality (\ref{TRIANGLE_TRICK}) to the expression within the bracket. This time vectors $\vec x$ and $\vec y$ have the following components:
\begin{eqnarray}
\vec x = (u_{kl},u_{pq}), \qquad \vec y = (v_{kl},v_{pq}).
\end{eqnarray}
The integrand is bounded by:
\begin{equation}
\sqrt{u_{kl}^2 + v_{kl}^2} + \sqrt{u_{pq}^2 + v_{pq}^2}
\ge \sqrt{(u_{kl} + u_{pq})^2 + (v_{kl} + v_{pq})^2}
\end{equation}
Let us consider the term involving vector $\vec u$ only. Since the lengths are positive
\begin{equation}
(u_{kl} + u_{pq})^2 \ge u_{kl}^2 + u_{pq}^2.
\end{equation}
Recall that $u_{kl}$ and $u_{pq}$ are projections onto orthogonal planes. One can introduce normal vectors to these planes, $\vec n_{kl}$ and $\vec n_{pq}$, respectively, and write
\begin{equation}
(\vec n_{kl} \cdot \vec u)^2 + u_{kl}^2 = 1,
\quad \textrm{and} \quad
(\vec n_{pq} \cdot \vec u)^2 + u_{pq}^2 = 1.
\label{ALMOST_THERE}
\end{equation}
Note that the scalar products are two components of vector $\vec u$ in the Cartesian frame build out of vectors $\vec n_{kl}$, $\vec n_{pq}$, and the one which is orthogonal to these two. Since vector $\vec u$ is normalized one has:
\begin{equation}
(\vec n_{kl} \cdot \vec u)^2 + (\vec n_{pq} \cdot \vec u)^2 \le 1,
\end{equation}
which implies for the sum of equations (\ref{ALMOST_THERE}):
\begin{equation}
u_{kl}^2 + u_{pq}^2 \ge 1.
\end{equation}
The same applies to vector $\vec v$ and one can conclude that
\begin{equation}
\sqrt{u_{kl}^2 + v_{kl}^2} + \sqrt{u_{pq}^2 + v_{pq}^2}
 \ge \sqrt{2}.
\label{PROJECTIONS_BOUND}
\end{equation}
Since the weight function $F(\vec u,\vec v)$ is normalized, the final inequality reads:
\begin{equation}
| E_{kl}(\varphi) + E_{k'l'}(0) |
+ | E_{pq}^{\perp}(\varphi) + E_{p'q'}^{\perp}(0)|
\le  4 - \frac{4}{\pi} |\sin\frac{\varphi}{2}|.
\end{equation}

\section{Appendix III: Corrigendum to the previous version on quant-ph (0704.2529v1)}

The experimental values of correlation functions 
were measured at the difference angle $\varphi = 20^{\circ}$ (on the Poincar{\' e} sphere) 
rather than $18.8^{\circ}$ as stated in the previous version. 
The reported violation of $3.2$ standard deviations refers to $\varphi = 20^{\circ}$.

In Appendix II the boundaries of all integrals over variables $\chi$ and $\psi$ should be the following: 
$\chi$ varies from $- 2 \pi$ to $2 \pi$; $\psi$ varies from $|\chi|/2$ to $2 \pi - |\chi|/2$.
Also, in equation (46) and following $F(\theta_u,\theta_v)$ should be replaced by
$\int_{-2 \pi}^{2 \pi} d \chi \int_{|\chi|/2}^{2 \pi - |\chi|/2} d \psi F(\theta_u,\theta_v,\psi,\chi)$.
\\

The final inequality of the paper and all its conclusions do not change.
The errors arose in the attempt for a concise presentation
but were never used in the actual computation.
In the present manuscript these errors are corrected.
We replaced Appendix II with a shorter derivation of the final inequality.
\\

We would like to thank Stephen Parrott for pointing us to the errors in the original paper.


\begin{thebibliography}{30}
\expandafter\ifx\csname natexlab\endcsname\relax\def\natexlab#1{#1}\fi
\expandafter\ifx\csname bibnamefont\endcsname\relax
  \def\bibnamefont#1{#1}\fi
\expandafter\ifx\csname bibfnamefont\endcsname\relax
  \def\bibfnamefont#1{#1}\fi
\expandafter\ifx\csname citenamefont\endcsname\relax
  \def\citenamefont#1{#1}\fi
\expandafter\ifx\csname url\endcsname\relax
  \def\url#1{\texttt{#1}}\fi
\expandafter\ifx\csname urlprefix\endcsname\relax\def\urlprefix{URL }\fi
\providecommand{\bibinfo}[2]{#2}
\providecommand{\eprint}[2][]{\url{#2}}

\bibitem[{\citenamefont{in}(2007)\citenamefont{published}}]{Groeblacher2007}
\bibinfo{author}{\bibfnamefont{This work was published}~\bibnamefont{in}}
  \bibinfo{journal}{Nature} \textbf{\bibinfo{volume}{446}},
  \bibinfo{pages}{871} (\bibinfo{year}{2007}).

\bibitem[{\citenamefont{Schr\"{o}dinger}(1935)}]{Schroedinger1935}
\bibinfo{author}{\bibfnamefont{E.}~\bibnamefont{Schr\"{o}dinger}},
  \bibinfo{journal}{Die\ Naturwissenschaften} \textbf{\bibinfo{volume}{48}},
  \bibinfo{pages}{808} (\bibinfo{year}{1935}).

\bibitem[{\citenamefont{Clauser et~al.}(1969)\citenamefont{Clauser, Horne,
  Shimony, and Holt}}]{CHSH1969}
\bibinfo{author}{\bibfnamefont{J.~F.} \bibnamefont{Clauser}},
  \bibinfo{author}{\bibfnamefont{M.~A.} \bibnamefont{Horne}},
  \bibinfo{author}{\bibfnamefont{A.}~\bibnamefont{Shimony}}, \bibnamefont{and}
  \bibinfo{author}{\bibfnamefont{R.~A.} \bibnamefont{Holt}},
  \bibinfo{journal}{Phys.\ Rev.\ Lett.} \textbf{\bibinfo{volume}{23}},
  \bibinfo{pages}{880} (\bibinfo{year}{1969}).

\bibitem[{\citenamefont{Leggett}(2003)}]{Leggett2003}
\bibinfo{author}{\bibfnamefont{A.~J.} \bibnamefont{Leggett}},
  \bibinfo{journal}{Found.\ of\ Phys.} \textbf{\bibinfo{volume}{33}},
  \bibinfo{pages}{1469} (\bibinfo{year}{2003}).

\bibitem[{\citenamefont{Einstein et~al.}(1935)\citenamefont{Einstein, Podolsky,
  and Rosen}}]{EPR1935}
\bibinfo{author}{\bibfnamefont{A.}~\bibnamefont{Einstein}},
  \bibinfo{author}{\bibfnamefont{B.}~\bibnamefont{Podolsky}}, \bibnamefont{and}
  \bibinfo{author}{\bibfnamefont{N.}~\bibnamefont{Rosen}},
  \bibinfo{journal}{Phys.\ Rev.} \textbf{\bibinfo{volume}{47}},
  \bibinfo{pages}{777} (\bibinfo{year}{1935}).

\bibitem[{\citenamefont{Einstein}(19 June 1935)}]{Einstein1935}
\bibinfo{author}{\bibfnamefont{A.}~\bibnamefont{Einstein}}
  \bibinfo{author}{\bibnamefont{to}~\bibnamefont{E. {S}chr\"odinger}},
  \bibinfo{howpublished}{Albert Einstein Archives, Jewish National and
  University Library, The Hebrew University of Jerusalem} (\bibinfo{year}{19
  June 1935}).

\bibitem[{\citenamefont{Bohr}(1949)}]{Bohr1949}
\bibinfo{author}{\bibfnamefont{N.}~\bibnamefont{Bohr}},
  \emph{\bibinfo{title}{Albert Einstein: Philosopher-Scientist}},
  vol.~\bibinfo{volume}{7} (\bibinfo{publisher}{Library of Living Philosophers,
  Evanston}, \bibinfo{year}{1949}).

\bibitem[{\citenamefont{Bell}(1964)}]{Bell1964}
\bibinfo{author}{\bibfnamefont{J.~S.} \bibnamefont{Bell}},
  \bibinfo{journal}{Physics} \textbf{\bibinfo{volume}{1}}, \bibinfo{pages}{195}
  (\bibinfo{year}{1964}).

\bibitem[{\citenamefont{Clauser and Horne}(1974)}]{CH1974}
\bibinfo{author}{\bibfnamefont{J.~F.} \bibnamefont{Clauser}} \bibnamefont{and}
  \bibinfo{author}{\bibfnamefont{M.~A.} \bibnamefont{Horne}},
  \bibinfo{journal}{Phys.\ Rev.\ D} \textbf{\bibinfo{volume}{10}},
  \bibinfo{pages}{526} (\bibinfo{year}{1974}).

\bibitem[{\citenamefont{Clauser}(2002)}]{Clauser2002}
\bibinfo{author}{\bibfnamefont{J.~F.} \bibnamefont{Clauser}},
  \emph{\bibinfo{title}{Quantum [Un]speakables:\ {F}rom {B}ell to {Q}uantum
  {I}nformation}} (\bibinfo{publisher}{R. A. Bertlmann and A. Zeilinger},
  \bibinfo{year}{2002}).

\bibitem[{\citenamefont{Greenberger et~al.}(1989)\citenamefont{Greenberger,
  Horne, and Zeilinger}}]{GHZ1989}
\bibinfo{author}{\bibfnamefont{D.~M.} \bibnamefont{Greenberger}},
  \bibinfo{author}{\bibfnamefont{M.}~\bibnamefont{Horne}}, \bibnamefont{and}
  \bibinfo{author}{\bibfnamefont{A.}~\bibnamefont{Zeilinger}},
  \emph{\bibinfo{title}{Bell's {T}heorem, {Q}uantum {T}heory, and {C}onceptions
  of the {U}niverse}} (\bibinfo{publisher}{Kluwer, Dordrecht},
  \bibinfo{year}{1989}).

\bibitem[{\citenamefont{Greenberger et~al.}(1990)\citenamefont{Greenberger,
  Horne, Shimony, and Zeilinger}}]{Greenberger1990}
\bibinfo{author}{\bibfnamefont{D.~M.} \bibnamefont{Greenberger}},
  \bibinfo{author}{\bibfnamefont{M.~A.} \bibnamefont{Horne}},
  \bibinfo{author}{\bibfnamefont{A.}~\bibnamefont{Shimony}}, \bibnamefont{and}
  \bibinfo{author}{\bibfnamefont{A.}~\bibnamefont{Zeilinger}},
  \bibinfo{journal}{Am.\ J.\ Phys.} \textbf{\bibinfo{volume}{58}},
  \bibinfo{pages}{1131} (\bibinfo{year}{1990}).

\bibitem[{\citenamefont{Freedman and Clauser}(1972)}]{Freedman1972}
\bibinfo{author}{\bibfnamefont{S.~J.} \bibnamefont{Freedman}} \bibnamefont{and}
  \bibinfo{author}{\bibfnamefont{J.~F.} \bibnamefont{Clauser}},
  \bibinfo{journal}{Phys.\ Rev.\ Lett.} \textbf{\bibinfo{volume}{28}},
  \bibinfo{pages}{938} (\bibinfo{year}{1972}).

\bibitem[{\citenamefont{Aspect et~al.}(1981)\citenamefont{Aspect, Grangier, and
  Roger}}]{Aspect1981}
\bibinfo{author}{\bibfnamefont{A.}~\bibnamefont{Aspect}},
  \bibinfo{author}{\bibfnamefont{P.}~\bibnamefont{Grangier}}, \bibnamefont{and}
  \bibinfo{author}{\bibfnamefont{G.}~\bibnamefont{Roger}},
  \bibinfo{journal}{Phys.\ Rev.\ Lett.} \textbf{\bibinfo{volume}{47}},
  \bibinfo{pages}{460} (\bibinfo{year}{1981}).

\bibitem[{\citenamefont{Aspect et~al.}(1982)\citenamefont{Aspect, Dalibard, and
  Roger}}]{Aspect1982}
\bibinfo{author}{\bibfnamefont{A.}~\bibnamefont{Aspect}},
  \bibinfo{author}{\bibfnamefont{J.}~\bibnamefont{Dalibard}}, \bibnamefont{and}
  \bibinfo{author}{\bibfnamefont{G.}~\bibnamefont{Roger}},
  \bibinfo{journal}{Phys.\ Rev.\ Lett.} \textbf{\bibinfo{volume}{49}},
  \bibinfo{pages}{1804} (\bibinfo{year}{1982}).

\bibitem[{\citenamefont{Weihs et~al.}(1998)\citenamefont{Weihs, Jennewein,
  Simon, Weinfurter, and Zeilinger}}]{Weihs1998}
\bibinfo{author}{\bibfnamefont{G.}~\bibnamefont{Weihs}},
  \bibinfo{author}{\bibfnamefont{T.}~\bibnamefont{Jennewein}},
  \bibinfo{author}{\bibfnamefont{C.}~\bibnamefont{Simon}},
  \bibinfo{author}{\bibfnamefont{H.}~\bibnamefont{Weinfurter}},
  \bibnamefont{and}
  \bibinfo{author}{\bibfnamefont{A.}~\bibnamefont{Zeilinger}},
  \bibinfo{journal}{Phys.\ Rev.\ Lett.} \textbf{\bibinfo{volume}{81}},
  \bibinfo{pages}{5039} (\bibinfo{year}{1998}).

\bibitem[{\citenamefont{Rowe et~al.}(2001)\citenamefont{Rowe, Kielpinski,
  Meyer, Sackett, Itano, Monroe, and Wineland}}]{Rowe2001}
\bibinfo{author}{\bibfnamefont{M.~A.} \bibnamefont{Rowe}},
  \bibinfo{author}{\bibfnamefont{D.}~\bibnamefont{Kielpinski}},
  \bibinfo{author}{\bibfnamefont{V.}~\bibnamefont{Meyer}},
  \bibinfo{author}{\bibfnamefont{C.~A.} \bibnamefont{Sackett}},
  \bibinfo{author}{\bibfnamefont{W.~M.} \bibnamefont{Itano}},
  \bibinfo{author}{\bibfnamefont{C.}~\bibnamefont{Monroe}}, \bibnamefont{and}
  \bibinfo{author}{\bibfnamefont{D.~J.} \bibnamefont{Wineland}},
  \bibinfo{journal}{Nature} \textbf{\bibinfo{volume}{409}},
  \bibinfo{pages}{791} (\bibinfo{year}{2001}).

\bibitem[{\citenamefont{Pan et~al.}(2000)\citenamefont{Pan, Bouwmeester,
  Daniell, Weinfurter, and Zeilinger}}]{Pan2000}
\bibinfo{author}{\bibfnamefont{J.-W.} \bibnamefont{Pan}},
  \bibinfo{author}{\bibfnamefont{D.}~\bibnamefont{Bouwmeester}},
  \bibinfo{author}{\bibfnamefont{M.}~\bibnamefont{Daniell}},
  \bibinfo{author}{\bibfnamefont{H.}~\bibnamefont{Weinfurter}},
  \bibnamefont{and}
  \bibinfo{author}{\bibfnamefont{A.}~\bibnamefont{Zeilinger}},
  \bibinfo{journal}{Nature} \textbf{\bibinfo{volume}{403}},
  \bibinfo{pages}{515} (\bibinfo{year}{2000}).

\bibitem[{\citenamefont{Zeilinger}(1986)}]{Zeilinger1986}
\bibinfo{author}{\bibfnamefont{A.}~\bibnamefont{Zeilinger}},
  \bibinfo{journal}{Phys.\ Lett.\ A} \textbf{\bibinfo{volume}{118}},
  \bibinfo{pages}{1} (\bibinfo{year}{1986}).

\bibitem[{\citenamefont{Aspect}(1999)}]{Aspect1999}
\bibinfo{author}{\bibfnamefont{A.}~\bibnamefont{Aspect}},
  \bibinfo{journal}{Nature} \textbf{\bibinfo{volume}{398}},
  \bibinfo{pages}{189} (\bibinfo{year}{1999}).

\bibitem[{\citenamefont{Grangier}(2001)}]{Grangier2001}
\bibinfo{author}{\bibfnamefont{P.}~\bibnamefont{Grangier}},
  \bibinfo{journal}{Nature} \textbf{\bibinfo{volume}{409}},
  \bibinfo{pages}{774} (\bibinfo{year}{2001}).

\bibitem[{\citenamefont{Born and Wolf}(1964)}]{Born1964}
\bibinfo{author}{\bibfnamefont{M.}~\bibnamefont{Born}} \bibnamefont{and}
  \bibinfo{author}{\bibfnamefont{E.}~\bibnamefont{Wolf}},
  \emph{\bibinfo{title}{Principles of {O}ptics:\ {E}lectromagnetic {T}heory of
  {P}ropagation, {I}nterference and {D}iffraction of {L}ight}}
  (\bibinfo{publisher}{Pergamon, Oxford}, \bibinfo{year}{1964}),
  \bibinfo{edition}{2nd} ed.

\bibitem[{\citenamefont{Bacon and Toner}(2003)}]{Bacon2003}
\bibinfo{author}{\bibfnamefont{D.}~\bibnamefont{Bacon}} \bibnamefont{and}
  \bibinfo{author}{\bibfnamefont{B.~F.} \bibnamefont{Toner}},
  \bibinfo{journal}{Phys.\ Rev.\ Lett.} \textbf{\bibinfo{volume}{90}},
  \bibinfo{pages}{157904} (\bibinfo{year}{2003}).

\bibitem[{\citenamefont{Ne'eman}(1986)}]{Ne'eman1986}
\bibinfo{author}{\bibfnamefont{Y.}~\bibnamefont{Ne'eman}},
  \bibinfo{journal}{Found.\ Phys.} \textbf{\bibinfo{volume}{16}},
  \bibinfo{pages}{361} (\bibinfo{year}{1986}).

\bibitem[{\citenamefont{Bohm}(1952)}]{Bohm1952}
\bibinfo{author}{\bibfnamefont{D.}~\bibnamefont{Bohm}},
  \bibinfo{journal}{Phys.\ Rev.} \textbf{\bibinfo{volume}{85}},
  \bibinfo{pages}{166} (\bibinfo{year}{1952}).

\bibitem[{\citenamefont{Dewdney et~al.}(1987)\citenamefont{Dewdney, Holland,
  and Kyprianidis}}]{Dewdney1987}
\bibinfo{author}{\bibfnamefont{C.}~\bibnamefont{Dewdney}},
  \bibinfo{author}{\bibfnamefont{P.~R.} \bibnamefont{Holland}},
  \bibnamefont{and}
  \bibinfo{author}{\bibfnamefont{A.}~\bibnamefont{Kyprianidis}},
  \bibinfo{journal}{J.\ Phys.\ A:\ Math.\ Gen.} \textbf{\bibinfo{volume}{20}},
  \bibinfo{pages}{4717} (\bibinfo{year}{1987}).

\bibitem[{\citenamefont{Wu and Shaknov}(1950)}]{Wu1950}
\bibinfo{author}{\bibfnamefont{C.~S.} \bibnamefont{Wu}} \bibnamefont{and}
  \bibinfo{author}{\bibfnamefont{I.}~\bibnamefont{Shaknov}},
  \bibinfo{journal}{Phys.\ Rev.} \textbf{\bibinfo{volume}{77}},
  \bibinfo{pages}{136} (\bibinfo{year}{1950}).

\bibitem[{\citenamefont{Kocher and Commins}(1967)}]{Kocher1967}
\bibinfo{author}{\bibfnamefont{C.~A.} \bibnamefont{Kocher}} \bibnamefont{and}
  \bibinfo{author}{\bibfnamefont{E.~D.} \bibnamefont{Commins}},
  \bibinfo{journal}{Phys.\ Rev.\ Lett.} \textbf{\bibinfo{volume}{18}},
  \bibinfo{pages}{575} (\bibinfo{year}{1967}).

\bibitem[{\citenamefont{Clauser and Shimony}(1978)}]{Clauser1978}
\bibinfo{author}{\bibfnamefont{J.~F.} \bibnamefont{Clauser}} \bibnamefont{and}
  \bibinfo{author}{\bibfnamefont{A.}~\bibnamefont{Shimony}},
  \bibinfo{journal}{Rep.\ Prog.\ Phys.} \textbf{\bibinfo{volume}{41}},
  \bibinfo{pages}{1881} (\bibinfo{year}{1978}).

\bibitem[{\citenamefont{Bell}(1985)}]{Bell1985}
\bibinfo{author}{\bibfnamefont{J.~S.} \bibnamefont{Bell}},
  \bibinfo{journal}{Dialectica} \textbf{\bibinfo{volume}{39}},
  \bibinfo{pages}{103} (\bibinfo{year}{1985}).

\end{thebibliography}
\end{document}